\documentclass[11pt]{article}
\usepackage{moriond,epsfig}

\def\Journal#1#2#3#4{{#1} {\bf #2}, #3 (#4)}

\def\NPB{{\em Nucl. Phys.} B}
\def\PLB{{\em Phys. Lett.}  B}

\def\PRD{{\em Phys. Rev.} D}

\def\eq#1{Eq.~(\ref{#1})}

\def\be{\begin{equation}}
\def\ee{\end{equation}}
\def\bea{\begin{eqnarray}}
\def\eea{\end{eqnarray}}

\begin{document}
\begin{flushright}
MAN/HEP/2008/7 \\
DFTT--9/2008
\end{flushright}
\vspace*{15mm}
\title{ANALYTICAL STUDIES FOR NON-PERTURBATIVE QCD OF JETS AT HADRON
COLLIDERS.}

\author{ M.~DASGUPTA$^a$, L. MAGNEA$^b$ and G. SALAM$^c$}
\vspace*{9mm}

\address{$^a$School of Physics and Astronomy, Oxford Road,\\
Manchester M13 3NP, England.\\
$^b$Dipartimento di Fisica Teorica, Universit{\`a} di Torino, and\\
INFN, Sezione di Torino, Via P. Giuria, I--10125 Torino, Italy.\\
$^c$LPTHE, CNRS UMR 7589; UPMC Univ.\ Paris 6;\\
Universit\'e Paris Diderot (Paris 7), 75252 Paris Cedex 05, France.}

\maketitle\abstracts{
Inspired by the success of analytical models for non-perturbative effects,
used to investigate event shape variables at LEP and HERA, we apply them
to a study of jets at hadron colliders such as the Tevatron and the LHC.
We find that simple analytical estimates are able to shed considerable light
on issues that could previously be tackled only through Monte-Carlo
simulations,
for example the role of different non-perturbative effects in various jet
algorithms.
In this context, we also provide testable numerical results for the
commonly studied
inclusive-jet $p_t$ distribution, and we introduce new observables that
could be
employed to verify our calculations.}

\section{Introduction}

With the LHC due to start later this year, there is considerable activity
geared
towards sharpening of theoretical and experimental tools, so as to
optimize its
discovery potential. A portion of this activity is directed at developing
a more
refined understanding of the physics of strong interactions (QCD), since they
will be ubiquitous at the LHC.  Since QCD has a non-perturbative aspect
that is
out of reach for the available tools of quantum field theory, there is an
immediate challenge to the level of precision that one may hope to achieve.
Inevitably, one has to deal at some level with the effects of parton
hadronization,
as well as with contamination from the non-perturbative underlying event that
accompanies the main hard process.

In an ideal world, one may for example envisage reconstructing clear mass
peaks --- or other kinematic structures ---
for some heavy decaying particle (for instance a SUSY particle, or a $Z'$
decaying
to jets at the LHC); in the real hadron collider environment, however,
these peaks
will be smeared by shifts and distortions in the energy spectrum of final
state jets,
induced by different QCD effects, so that the signal may even be
altogether washed
out. The smearing effects will involve both initial and final state QCD
radiation, as
well as non-perturbative energy flows arising from hadronization and the
underlying
event. To minimise such smearing requires some understanding of the
dependence
of each effect on the experimental parameters involved in the study, in
particular on
the choice of jet-algorithm and on the choice of jet size (which is
governed by a
``radius'' parameter $R$). While perturbative contributions can be obtained
using Feynman graph techniques, it is less clear how to acquire
information on
non-perturbative effects. This is the question that we shall focus on
below: we will
employ analytical models~\cite{DW1} that have been very successful in the
context
of DIS and $e^+ e^-$ event shape studies to the more complex environment of
hadron collisions.

\section{Non-perturbative tools for jet physics}

The toolkit for non-perturbative (NP) physics of QCD jets has been thus far
rather limited, comprising almost exclusively Monte Carlo (MC) studies using
mostly {\tt HERWIG} and {\tt PYTHIA}. While MC's are indispensable tools
in this
and other regards, they have their own shortcomings, and a certain amount of
analytical insight is thus, in our opinion, a welcome addition.  For
example, it is
not straightforward to gain information from MC studies on the functional
dependence
of NP corrections on jet parameters such as radius, flavour and $p_t$,
while this
information is provided immediately by the analytical estimates we will
derive.
The lack of parametric information, in turn, gives rise to a lore of
qualitative statements
that may or may not be supported by a quantitative analysis. One may hear, for example,
that the $k_t$ algorithm~\cite{KT} suffers more significantly from
underlying event (UE) contamination, as compared to cone algorithms~\cite{CONE},
which are supposed to be more significantly affected by hadronisation. We
find
that, if one chooses the same value of jet radius in either case, there
are no differences
between algorithms in a first-order calculation. For the UE,
calculated to the next
order~\cite{CSS}, one sees as much
variation between different cone algorithms as between cones and the
$k_t$ algorithm.\vspace{-0.5em}

\subsection{The Dokshitzer-Webber model applied to jets}

We shall first examine, as an example, hadronization corrections to a jet
transverse
momentum $p_t$, and then turn to the underlying event contribution.
To obtain our main analytical results for hadronization corrections, it is
sufficient to
use the renormalon-inspired model developed by Dokshitzer and
Webber~\cite{DW1}
(DW). This model has been widely used for QCD studies at HERA and LEP, and
has
been followed by several theoretical developments~\cite{DMW,DW2,MIL,KS,G},
which
have firmly established its physical features in the context of our
understanding
of perturbative QCD. To understand our central result, it is however
sufficient to
use the model in its original form. In the DW model, hadronization is
associated to
the emission of a soft gluon with transverse momentum $k_t \sim \Lambda_{
\mathrm{QCD}}$ (``gluer''). While the strong coupling associated to such an
emission, $\alpha_s (k_t)$, is divergent within perturbation theory, one
assumes
that it can be replaced, in the infrared, by a physically meaningful
infrared finite and
universal coupling. One then calculates the change $\delta p_t$ in the
transverse
momentum of a jet due to gluer emission, and one averages this change over
the
gluer emission probability.

In general the calculation will depend on the details of the hard process
of which
the triggered jet is a part.  A full calculation in the threshold limit of
hadronic dijet
production has been reported in Ref.~\cite{DMS}. The calculation there
reveals
that the hadronization contribution is singular in the jet radius $R$, as
$R \to 0$,
{\it i.e.} in the limit of narrow jets. This most significant feature is
in fact universal,
and applies to jet production in any hard process; moreover, the leading
behavior
in $R$ can be derived with a simple calculation, as we illustrate below.

Consider the emission of a soft gluon from a hard parton (say a quark to be
definite), such that the gluon is not recombined with the quark jet. We
will work
in the collinear approximation, which is sufficient to reproduce the leading
small-$R$ behaviour. If the transverse momentum of the quark jet was $p_t$
before gluon emission, it becomes $z p_t$ after the emission, with $z$
the fraction of the initial quark momentum carried by the final quark, so
that
in the soft limit $z \to 1$. The change in $p_t$ induced by gluon emission is
then $\delta p_t  = ( z - 1 ) p_t$. Averaging this over phase space with the
appropriate probability distribution leads to \footnote{The condition that
the gluon
not be recombined with the jet reduces to $\theta > R$ in the soft limit
for all
the commonly used jet algorithms.}
\begin{equation}
\left \langle \delta p_t \right \rangle = p_t \int \frac{d
\theta^2}{\theta^2}
\int d z \, (z - 1) \, P_{q q} (z) \, \frac{\alpha_s \left(\theta z (1 -
z) p_t
\right)}{2 \pi} \, \Theta \left( \theta - R \right) \, .
\label{colleq}
\end{equation}
\begin{figure}
\begin{center}
\epsfig{file=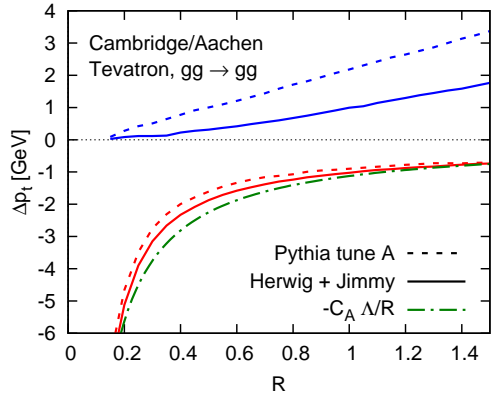,angle=270, width =0.4\textwidth}
\caption{Hadronisation (negative) and underlying event (positive)
contributions to
jet $p_t$, as functions of the jet radius $R$, for gluon jets at the
Tevatron.
\label{fig:dptvR}}
\end{center}
\end{figure}
\noindent In the perturbative regime \eq{colleq} gives a $\log R$
behaviour, which is
a reflection of the collinear enhancement. To evaluate non-perturbative
contributions we change variable to $k_t = z (1 - z) \theta p_t$, we
insert the soft limit of the splitting function $P_{q q } = 2/(1 - z)$,  and
we substitute to the coupling its non-perturbative modification $\delta
\alpha_s$, corresponding to `gluer' emission. We then integrate over
$\theta$ and $z$, which gives
\begin{equation}
\left \langle \delta p_t \right \rangle^{h} = - \frac{4}{R} \, C_F \, \int
\frac{dk_t}{k_t} k_t \frac{\delta \alpha_s (k_t)}{2 \pi} \, ,
\label{deltapt}
\end{equation}
where $\delta \alpha_s$ is the non-perturbative QCD coupling minus its
perturbative counterpart, and it is non-vanishing only in the infrared
region,
$0 < k_t < \mu_I$, with $\mu_I$ an infrared matching scale conventionally
taken to be $\mu_I = 2$ GeV. The value of the integral of $\delta \alpha_s
(k_t)$ cannot be computed, but it can be extracted from event shape
variables,
under the assumption of universality. We arrive then at a simple result
for the
$p_t$ shift of a quark jet, which amounts to $\approx - 0.5/R$ GeV. For a
gluon jet the corresponding result is obtained by replacing $C_F$ with $C_A$
in \eq{deltapt}.

The behaviour of underlying event contributions to the same observable, on
the other hand, is regular, and vanishes like $R^2$ as $R \to 0$, in stark
contrast with \eq{deltapt}. This result is natural since the underlying event
is disentangled from the dynamics of the jet, which serves merely as a
receptacle for soft radiation from partons uncorrelated with the hard
scattering.
Assuming a uniform rapidity distribution for the soft radiation gives a
contribution to $\delta p_t$ proportional to the jet area~\cite{CSS}, with a
functional dependence on $R$ given by $R J_1(R) = R^2 + {\mathcal{O}}(R^4)$.

We have compared our expectations for the $R$ dependence with Monte
Carlo event generators, and the results are shown in Fig.~\ref{fig:dptvR}.
One observes that the $1/R$ hadronization correction is in good agreement
with the event generators {\tt HERWIG} and {\tt PYTHIA}, in both shape
and normalization, over virtually the full range of $R$ studied.
In contrast, while the underlying event varies with $R$ as expected, its
normalisation is different depending on the event generator model. We also
emphasize that very similar results are obtained with all commonly used jet
algorithms, so that we have displayed just the Cambridge/Aachen~\cite{CA}
algorithm. We conclude that by varying $R$ it is possible to enhance or
reduce the sensitivity to one non-perturbative effect or the other, as
desired,
which leads to the possibility of isolating and testing individually the
different
sources of non-perturbative contributions to jets at hadron colliders.
We note finally that the size of the underlying event contribution, unlike
that
of hadronisation, is not under theoretical control, and is different for
{\tt HERWIG} and {\tt PYTHIA} at Tevatron energies. Further work is needed
to obtain a less ambiguous picture for this component of NP physics.

\section{Experimental tests and applications}

\begin{figure}
\begin{center}
\epsfig{file=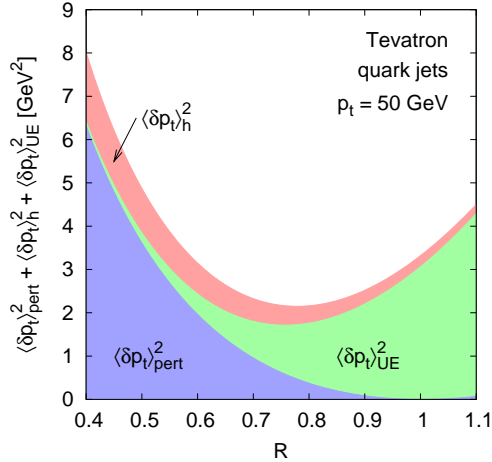,width =0.4\textwidth}
\caption{The dispersion of jet $p_t$ as a function of jet radius, due to
perturbative and non-perturbative QCD effects, for 50 GeV quark jets at
the Tevatron. The minimum value for the total dispersion corresponds
to the best value of $R$ if one wishes to minimize all QCD effects.
\label{fig:bestr}}
\end{center}
\end{figure}
We briefly present here some experimental avenues to corroborate and
exploit the results mentioned above. A fuller account is available in our
article~\cite{DMS}. One idea that emerges from computing the different
$R$ dependencies of perturbative and non-perturbative QCD effects is
that of optimal values of $R$ for studies involving jets. In the sort of
study
we mentioned before, aiming at the reconstruction of the mass of a heavy
particle decaying to jets, we would like to minimise the dispersion on jet
$p_t$ due to all QCD effects (perturbative and non-perturbative). A detailed
study of this dispersion would require a knowledge of correlations between
different physical effects, which is not available with current tools. To
get a
qualitative understanding, one may approximate the true dispersion with the
uncorrelated sum
\begin{equation}
\langle \delta p_t^2 \rangle  = \langle \delta p_t \rangle ^2_h
 + \langle \delta p_t \rangle ^2_{\mathrm{UE}}  +
\langle \delta p_t \rangle ^2_{\mathrm{PT}} \, .
\label{uncorsum}
\end{equation}
Each term in the sum has a characteristic $R$ dependence at small $R$,
with the perturbative piece varying as $\log R$ , the hadronisation
correction
as $- 1/R$, and the underlying event contribution as $R J_1 (R)$. The result
is plotted for 50 GeV quark jets at the Tevatron in Fig.~\ref{fig:bestr},
where
we displayed the dispersion due to each effect separately, as well as the
approximate
total dispersion, whose minimum corresponds to the optimal $R$.

While Fig.~\ref{fig:bestr} reflects what could be achieved with current
knowledge
about the $R$ dependence, it should not be taken too literally as far as
the precise
value of the optimal $R$ is concerned, since we neglected correlations, and
furthermore we have oversimplified the perturbative piece, retaining only
the leading
small $R$ behaviour. The general features of Fig.~\ref{fig:bestr} are
however robust,
since they follow from the different parametric dependence on $R$ of the
various
physical effects. From our studies we are also able to predict how the
optimal $R$
may change with a change of jet parameters such as ``flavour'' or $p_t$.
As might
be expected, a gluon jet favours a larger $R$ value than a quark jet, and
likewise the
optimal $R$ rises in a predictable manner with increasing jet $p_t$
(see~\cite{DMS}
for details).

For QCD studies, involving, say, the determination of $\alpha_s$ from jet
observables, one may again search for an optimal $R$: in this case however
one
should seek to minimize only the non-perturbative contributions. One
finds~\cite{DMS}
that the optimal $R$, in this case, is proportional to the cube root of
the ratio of the
characteristic scales for hadronization and underlying event.

Various direct experimental tests can be carried out to check our
predictions.
In this regard one may for example study inclusive jets at HERA, where the
steeply
falling $p_t$ spectrum would be approximately shifted by the $1/R$
hadronisation
effect. Hence a study of inclusive jets with variable $R$ would provide a
valuable
opportunity to confirm our results. Similarly studies at the Tevatron
could lead to
a direct determination from data of the scale of the underlying event,
addressing the
current disagreement between the MC models of {\tt HERWIG} and {\tt
PYTHIA}. It
is also possible to define operationally, and measure directly as a
function of $R$,
the change in the jet $p_t$ due to nonperturbative effects as one changes
the jet
algorithm or the jet parameters; this definition can be implemented
in Monte Carlo studies and could be useful to determine the non-perturbative
scales associated with hadronization and underlying event.

To conclude, we would like to emphasise the role of simple analytical
studies, which
are however well grounded in the technology of perturbative QCD, in order to
obtain information about complex non-perturbative properties of jets. This
information, reflected for example in the dependence on the jet radius of
various jet
observables, ought to be of use in carrying out precision studies
involving jets at
current and future colliders. We would especially like to emphasize the
importance
of maintaining flexibility in the choice of jet algorithm and jet
parameters, since
our results show that choices that may be very useful for one class of
studies
may lead to poor results for other cases.
\vspace{-1em}

\section*{Acknowledgements}
\label{sec:acknolwedgments}

Work supported by MIUR under contract 2006020509$\_$004, by the French
ANR under contract ANR-05-JCJC-0046-01 and by the European Community's
Marie-Curie Research and Training Network `HEPTOOLS', under contract
MRTN-CT-2006-035505.

\end{document}